\documentclass[aps, prl, reprint]{revtex4-2}
\usepackage[T1]{fontenc}
\usepackage{geometry}
\usepackage{graphicx}
\usepackage{float}
\usepackage{chemformula} % Formulas using \ch{}
\usepackage[version=4]{mhchem} % Formulas using \ce{}
\usepackage{xcolor}
\usepackage{amsmath}
\usepackage{comment}
\usepackage{hyperref}

\newcommand{\vm}{\vec{M}}
\newcommand{\vmm}{\vec{m}}

\newcommand{\vE}{\vec{E}}
\newcommand{\vP}{\vec{P}}
\newcommand{\vD}{\vec{D}}

\newcommand{\vv}{\vec{v}}
\newcommand{\vs}{\vec{\sigma}}
\newcommand{\vL}{\vec{\Lambda}}
\newcommand{\vC}{\vec{C}}

\date{\today}
\begin{document}

\title{The geometry of the CISS effect}

\author{P. Hedegård}
\email{hedegard@nbi.ku.dk}
\affiliation{Niels Bohr Institute, University of Copenhagen, Denmark}

\author{A. Kazimir}
\author{C. Lamers}
\affiliation{Institute for Drug Discovery, Faculty of Medicine, Leipzig University, Germany}

\author{L.T. Baczewski}
\affiliation{Institute of Physics, Polish Academy of Sciences, Warszawa, Poland}

\author{T.N.H. Nguyen}
\affiliation{Institute of Physics, Chemnitz University of Technology, Germany}

\author{C. Tegenkamp}
\email{christoph.tegenkamp@physik.tu-chemnitz.de}
\affiliation{Institute of Physics, Chemnitz University of Technology, Germany}

\begin{abstract}
Using scanning tunneling microscopy, and a careful selection of chiral molecules and anchoring groups, we systematically carried out a series of magnetoresistance experiments. We observed a reversal of the signal upon changing the magnetization direction, the molecular chirality, and the molecular orientation. The orientation of the molecules and the magnetization of the substrate are vectors, and by associating an axial vector with the molecule, the observations can be explained. Other recent experiments, among them null experiments showing no CISS related magnetoresistance can be explained using our framework. The physical interpretation is, that the polar electrical polarization of the molecule cannot play a direct role, while the axial magnetic polarization vector operator is a much more realistic candidate. This vector plays an important role in the creation of a magnetic moment in the interface between molecule and metallic lead. 
\end{abstract}

\maketitle

%\section{Introduction}
\noindent\textit{Introduction --} The chirality-induced spin selectivity (CISS) phenomenon, first observed in helical molecules, is remarkably robust and manifests in a variety of forms \cite{Ray1999, Mishra2013, Naaman2019, Gohler2011, Eckvahl2023, Bloom2024, Rana2025}.

One of the most extensively studied systems is the magnetoresistance (MR) of currents flowing between two metallic leads, one of which is magnetized, separated by a layer of chiral molecules \cite{Yang2019, Tirion2024, Mondal2021, Ha2023}. The current value is compared between two measurements for the opposite magnetization direction of a magnetic layer. 
Under otherwise identical conditions, one would normally expect the two experiments to yield the same current.  
But, it is important, that the detailed atomic geometry of the hybrid interface  is unchanged; all atoms should be in the same place and the charge distribution should not change when magnetization is reversed. 
In this idealized two-terminal setup, one can rigorously show that, to linear order in the applied bias voltage, the current is independent of the magnetization direction $\vm$ as long as time-reversal symmetry (TRS) is preserved. This is a consequence of the famous Onsager reciprocity relation which reads for the current $I$:
\begin{equation}
    G(\vm,g)V = G(-\vm,g)V.
\end{equation}
Here $g$ refers to all other parameters such as the geometry. 
In the absence of electron correlations, TRS imposes constraints even in the nonlinear regime. This is evident from the Büttiker–Landauer formula for the electrical current in a two-terminal setup with metallic leads \cite{Buttiker1986, Buttiker1988}.

In this context, the CISS effect causes that the currents in the two situations are {\em not} identical and the (unnormalized) CISS-MR signal is given by: 
\begin{equation}
    \Delta I(V,\vm,g) = I(V,\vm,g) - I(V,-\vm,g).
\end{equation}

In addition to the magnetization direction, other parameters have been varied in such experiments influencing also the CISS effect, e.g., (i) the chirality of the molecules, (ii) the orientation of the molecules, (iii) 
racemic or non-racemic mixtures and (iv) ordered or random orientation of molecules. In case of polypeptides, some of these experiments have been performed \cite{Kiran2016, Ha2022, Ha2023, Ha2024}. 
The first two cases involve single molecules, whereas the last two correspond to measurements averaged over many molecules. Even in the single- or few-molecule experiments, such as those based on scanning tunneling microscopy (STM) or break junctions, many realizations of the junction must be measured to obtain a reliable CISS signal. 

Several relevant physical quantities are vector quantities. Certainly, the magnetization $\vm$ is a vector. If one expands the CISS-MR signal to lowest order in $\vm$ one gets
\begin{equation}
    \Delta I(V,\vm,g) \approx \vC \cdot \vm,\quad \vC = \left.\frac{\partial I}{\partial\vm}\right|_{\vm=0}. 
    \label{eqn3}
\end{equation}

The question arises, what is the vector $\vC$? The purpose of this work is to provide a partial answer to this question. By performing a set of in total 10 experiments, where the geometry of the setup is systematically varied. This involves changing the chirality and orientation of the molecule as well as the magnetization of the substrate. 
Based on these findings, we can establish phenomenological rules that $\vC$ should satisfy. Concretely, our observations can be consistently explained by the emergence of a magnetization at the hybrid interface induced by the helical molecule.

%\section{Methods}
\noindent\textit{Methods --} The CISS-MR measurements were performed using ambient scanning tunneling microscopy (STM) and spectroscopy (STS). STM tips were made from 0.25~mm Au wire, and all measurements were carried out at 300~K. To quantify the spin polarization of transmitted electrons, I--V spectra were recorded at setpoints $V_b = 0.5$~V and $I_t = 100$~pA. Each spectrum represents the average of at least ten measurements.
Alanine-rich helical peptides (also named as polyalanine (PA) here) containing 16 amino acid residues and consisting of alanine (A), lysine (K) and cysteine (C or Cys) , C[AAAAK]$_3$, with a total length of 2.4~nm, were used as the helical molecular system. The peptides were obtained from Sigma Aldrich (right handed L-, Cys\textsubscript{N}-, left-handed D-, Cys\textsubscript{N}-) and right-handed (or L-, Cys\textsubscript{C}- ) [AAAAK]$_3$C was synthesized using standard Fmoc-based solid-phase peptide synthesis (SPPS) protocols. The purity of the final products was approximately 
90\%, as determined by analytical RP-HPLC (see Supplement Material \cite{SM}).
The peptides were prepared with either  the C- or N-terminal cysteine (Cys$_C$ or Cys$_N$) to control adsorption and orientation of the helices relatively to the magnetic substrate and STM tip.
The PA helical molecules of a different helical chirality  (right-handed consisting of L-amino acids, left-handed consisting of D-amino acids) and with Cys$_C$ or Cys$_N$ were deposited on MBE-grown epitaxial magnetic nanostructures consisting of Al$_2$O$_3$/Pt/Au(20~nm)/Co(1.2~nm)/Au(5~nm). The Co layer exhibits out-of-plane magnetization with a coercive field of 16~mT, easily switchable by an external magnetic field higher than the coercive field.
Furthermore, also STM tips functionalized with Cys$_N$ PA were  used to reorient the molecules as previously shown in Ref.~\cite{Ha2024}. Thereby, the tips were prepared by immersing freshly cut Au wires into 0.3~mM molecular solutions.
Further experimental details are given in Refs.~\cite{Ha2022,Ha2024} and in the Supplemental Material \cite{SM}.\\

%\section{Results and discussion}

\begin{figure}
                    \centering
            \includegraphics[width=0.62\linewidth]{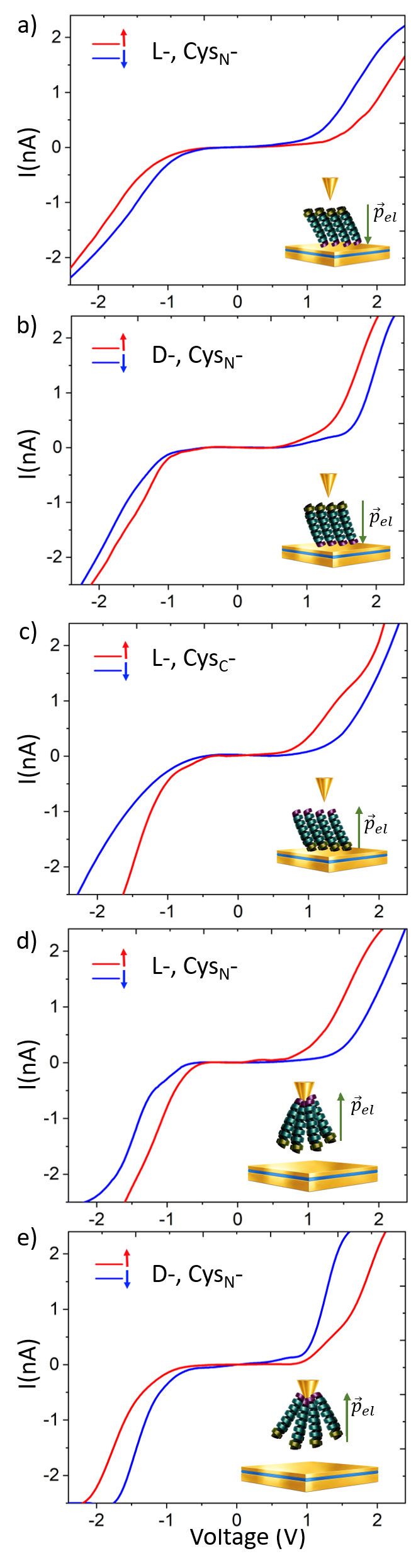}
        
    \caption{IV-curves for 16-mer PA molecules taken for the two out-of-plane magnetization directions ($\vm$=$\uparrow, \downarrow$) of the Co-layer with of GMR hybrid set up (a , b, and c) versus the TMR hybrid set up (d, e). The insets sketches the assembly of the (L,D)- and (Cys$_C$,Cys$_N$)-PA molecules on the  Au/Co/Au surface (a-c) and  Au tip (d,e). The orientation of the PA molecules is associated with electric dipole vector $\vP$.}
    \label{FIG1}
\end{figure}

%\section{Results and discussion:}
\noindent\textit{Results and discussion --} 
The IV-curves obtained for different orientations and chirality of the molecules with respect to the sample surface and STM tip as well as different magnetization directions are shown in Fig.~\ref{FIG1}. Thereby, the overall shape of the IV-curves are rather similar for all configurations. As seen in Fig.~\ref{FIG1}(a), the adsorption of the chiral molecule on the Au-surface reveal different tunneling currents $I$ depending on the two out-of-plane directions of the Co-layer magnetization $\vm$($\uparrow$,$\downarrow$). The normalized difference of the currents allows to define the  CISS-MR value as a function of the bias voltage \cite{Ha2024}. The CISS-MR can be seen best at higher bias voltages in our ambient temperature measurements. Reference measurements performed at low temperatures  under ultra high vacuum conditions reveal also for low bias voltages a difference in MR, thus the CISS for the PA shows up in the linear regime. For details see the Supplemental Material \cite{SM}.

From the IV-curves, we can extract the sign of the CISS-MR, i.e., the sign of $\vC\cdot\vm$ as introduced in Eq.~\ref{eqn3}. In Tab.~\ref{summarythiswork}, we summarize our findings. We denote $\vC$ by an $\uparrow$ when $\vC$ is aligned with $\vm$, and by a $\downarrow$ when $\vC$ is opposite to $\vm$ each for the high electron transmission. The molecular orientation is also indicated by arrows, which are parallel to the molecules’ electric dipole moment. Most importantly, the vector $\vC$ does not follow the electric dipole vector orientation. For example, when the molecular orientation and the electric dipole vector are kept fixed while the chirality is changed, the vector $\vC$ is reversed.  

\begin{table}[t]
    \centering
    \begin{tabular}{|c|c|c|p{1cm}|} \hline
      Chirality   & Orientation, $\vP$ & Anchor site& \hfil$\vC$\hfil \\ \hline \hline
        L & $\downarrow$ & substrate & \hfil$\downarrow$\hfil \\
        D & $\downarrow$ & substrate & \hfil$\uparrow$\hfil  \\
        L & $\uparrow$ & substrate & \hfil$\uparrow$\hfil \\
        L & $\uparrow$ & tip & \hfil$\uparrow$\hfil \\
        D & $\uparrow$ & tip & \hfil$\downarrow$\hfil \\ \hline
    \end{tabular}
    \caption{Summary of the experimental results of this study for L- and D-PA molecules. The orientation and binding site of the PA molecules are defined by the direction of the electric dipole moment $\vP$ and the anchor site, respectively. The orientation of $\vC$ is defined by the direction of the magnetization $\vm$ of the Co layer for which high transmission was observed.}
    \label{summarythiswork}
\end{table}

\begin{table}[htbp]
    \centering
    \begin{tabular}{|c|c|p{1cm}|} \hline
      Chirality   & Orientation $\vP$&  \hfil$\vC$\hfil \\ \hline \hline
        L & $\uparrow$ &  \hfil$\downarrow$\hfil \\
        D & $\uparrow$ & \hfil$\uparrow$\hfil  \\
        L & $\downarrow$ & \hfil$\uparrow$\hfil \\
        D & $\downarrow$ & \hfil$\downarrow$\hfil \\ \hline
    \end{tabular}
    \caption{Summary of the experimental results of Aragonès et al.\cite{Aragones2025}. The symbols are defined as in Tab.~\ref{summarythiswork}.}
    \label{summaryaragones}
\end{table}

In the recent work by Aragonés et al.\cite{Aragones2025}, the CISS-MR has been measured using an STM break junction technique. As molecules also $\alpha$-helical peptides with a fixed sequence of amino acids with C and N terminals were used. Instead of physically changing the orientation of the molecule, they instead use molecules with the retro-sequence. From a physical point of view, this is quite close to molecules which has been rotated. From their reported values of the conductance $G$, we  determined the CISS-MR and the direction of $\vC$, which is summarized in Tab.~\ref{summaryaragones}.

For both sets of experiments, we observe that the $\vC$ vector is transforming as an axial vector, i.e.~it does change orientation when going from right-handed to left-handed molecules.

By definition, $\vC$, is a vector quantity. It is defined as the gradient of a scalar quantity, the current differentiated with respect to $\vm$. The magnetization is an axial vector, hence we should expect that $\vC$ is also an axial vector. If is was not, the current necessarily had to be a pseudo scalar, i.e.~it would change sign, if the physical setup was a mirror image of the original setup, with the mirror plane containing the molecule and mirroring the leads onto themselves. 

What are possible physical candidates for the observed $\vC$ vector?
In order to answer this question, we invoke the Onsager reciprocity principle. According to this the conductance should be the same if all magnetic fields and magnetizations are reversed. This means, that the $\vC$ vector will have to change sign when all magnetizations are reversed. The simplest interpretation is that $\vC$ is proportional to some magnetization different from $\vm$, i.e. $\vC = \alpha \vmm$. 
The precise location of the molecule-induced magnetization has not yet been established. It could be in the Au layer close to molecule. This has been postulated by many, and in a recent experiment, Baljozovi\'c et al.\cite{Baljozovic2026} actually measures such an induced moment, and provides a credible physical mechanism that can produce such a moment. It could also reside as a small magnetic domain in the Co layer. This is not likely, since the strong ferromagnetic exchange coupling in the magnet would prevent that. In either case, electrons has to pass this magnetized region before it enters the magnet, and there will be a contribution to the conductance proportional to $\vmm\cdot\vm$, as known from the spintronic effects of giant magnetoresistance (GMR) and tunnel magnetoresistance (TMR) \cite{Schuhl2005}.

The reversal of the CISS-MR by chemically flipping the molecule is not new and has been reported previously~\cite{EckshtainLevi2016,Clever2022}, although it received little attention at the time.
Recently, the importance of the precise geometrical orientation of the chiral molecules has in fact been reported by Li et al.\cite{Latha2025}. Their finding indirectly supports the main conclusion of this work: namely, that a chiral molecule is associated with an axial vector, and that the orientation of this vector determines the magnetoresistance. Li et al. have studied a number of different chiral molecules using an STM break-junction technique. Interestingly, they did not report a CISS effect, although the setup is quite similar to that used in our work. But, the most important difference, is that the molecules in their study ((1R)-4,4'-bis(4-methylthio)phenyl)-2,2'dimethoxy-1,1'-binaphtalene) has thiol groups at both ends. Hence, one expects that it is random which end of the molecule attaches to the Au layer and which end attaches to the STM. This means that the associated $\vC$ vector for this molecule will have a random sign, and upon averaging over many molecules the CISS effect will cancel. So, these experiments constitute an important support of our findings.   

A final set of experiments that elucidates the importance of the molecule orientation are performed on superparamagnetic nanoparticles. Koplovitz et al.\cite{Koplovitz2019}, has reported experiments where helical 36-mer PA molecules  (C[AAAAK]\textsubscript{7}-COOH) quite similar to the ones used in this work, were deposited on superparamagnetic iron oxide nanoparticles (SPION) 10 nm in size. As in our case the orientation of the molecules were controlled by having thiols at one end and a carboxylic group at the opposite end. The thiol end is the one binding to a Ag surface. The carboxylic end binds preferentially to the SPION. In one experiment, SPIONs are added to a layer of oriented AHPAs. In that case the SPION induces a magnetization with a direction dictated by the handedness of the AHPA. In a control experiment, AHPAs are attached to the SPIONs from all sides, and no magnetization is observed. 

\vspace{1ex}
In the following we will formulate a single hypothesis, that  explains all the findings of our work and the results discussed above: A chiral molecule will generate a magnetization, $\vmm$, in the immediate neighborhood of the molecule. That magnetization has a direction, which is a vector, thus directly associated with the geometrical orientation of the molecule. So, if the molecule is rotated or mirrored, it will transform as an axial vector. 

Such a vector does in fact exist in a chiral molecule. It is the magnetic dipole vector associated with spin-orbit coupling (SOC). The SOC is in general given by $H_{SOC} = \vL\cdot\vs$. It has it's origin in the Dirac equation, where $\vL$ is of the form 
\begin{equation}
    \vL = -\frac{e\hbar}{2mc^2}(\vE\times\vv) 
\end{equation}
where $\vE$ is the electrical field felt by the electron and $\vv$ its velocity. In order to have a large SOC, one should look for fast electrons in a strong electrical field. Those are found in the atoms. In atoms (or in atomic orbitals) where the electrical field is radial, and the velocity is perpendicular to the radius vector, the $\vL$ becomes proportional to the angular momentum $\vL  \propto \vec{L}$. $\vL$  is an axial vector, since it is a sum of axial vectors. This in turn is proportional to the orbital magnetic dipole moment of the electron. Already London showed in 1937  that for a molecule or a solid with many atoms, the magnetic dipole moment operator can be written as a sum over operators belonging to the individual atoms: $\vec{\mu} = \sum_i \gamma_i \vec{L}_i$ \cite{london1937theorie,ditchfield1974self,wolinski1990efficient}. In fact, this operator is fundamental both for understanding circular dichroism (CD) and the SOC in molecules and probably the reason why CISS-MR and CD spectroscopy show strong similarities \cite{Amsallem2023}.  CD is a standard technique for probing the chirality of molecular systems. A central feature of all experiments probing the chirality  is the ability to distinguish between left- and right-handed enantiomers, a capability that is also demonstrated in the experiments presented in this work. Obviously, $\vL$ is an operator, so for different experiments it corresponds to an appropriate expectation value or ensemble average, which we denote as $\vD=tr(A\vL)$,
of this operator being measured.

Our suggestion is, that in proximity to the chiral molecule a magnetization, $\vmm$ is induced. Due to some not yet understood mechanism, this magnetization is proportional to the vector $\vD$ associated with the SOC of the chiral molecule. If the molecule changes it's orientation, so does the $\vD$ and $\vmm$. Further, if a left-handed molecule is replaced by it's right-handed counterpart, the vectors $\vD$ and $\vmm$ are changed accordingly.

\begin{figure}[t]
    \centering
    \includegraphics[width=\linewidth]{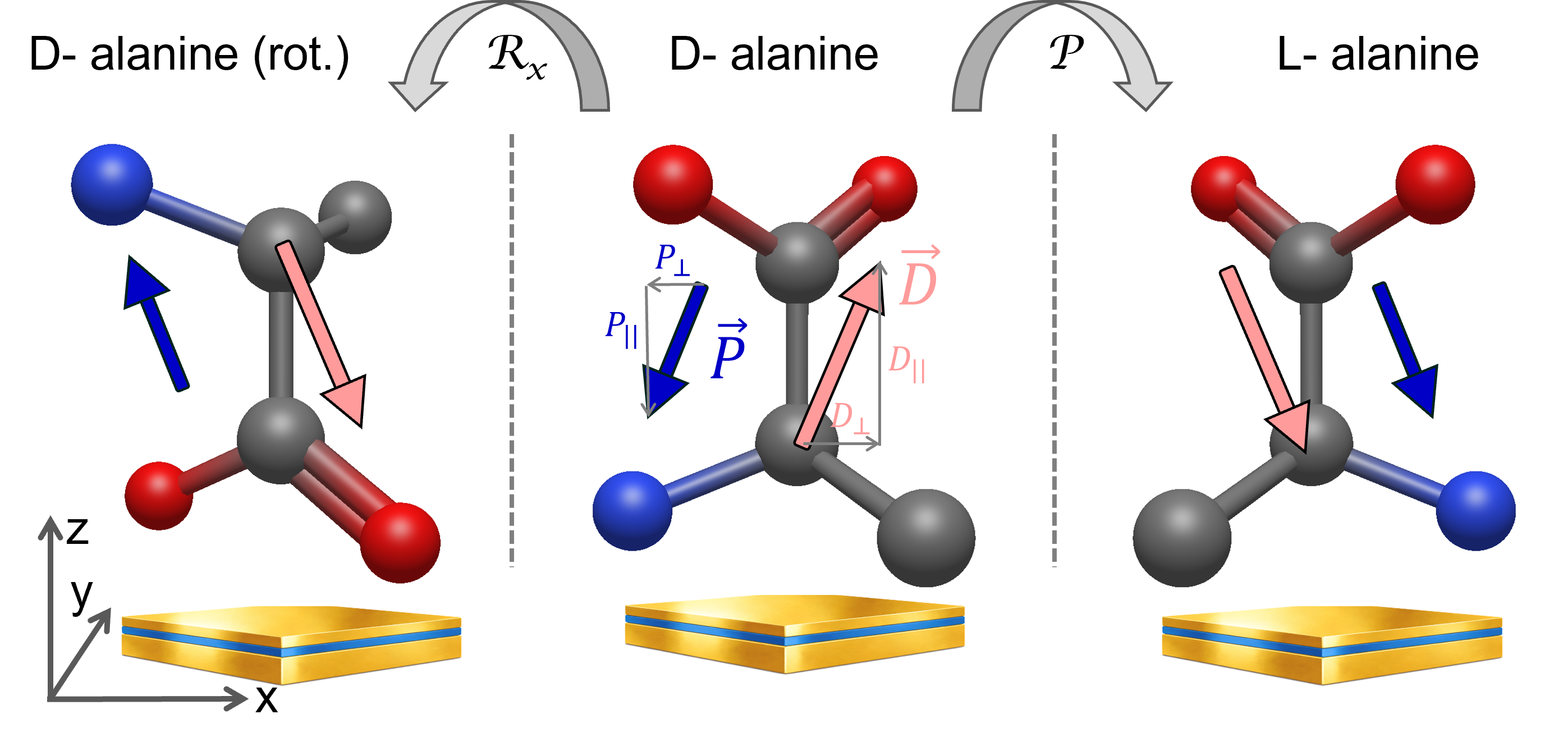}
    \caption{The parity operation (right) and rotation (left) of D-alanine in its zwitterionic form, serving as a building block for the PA molecules shown in Fig.~\ref{FIG1}. The electric dipole vector $\vP$\ and molecular orbital vector $\vD$  as well as their different transformation behaviors are shown too. For the sake of simplicity, the vectors are plotted in the xz-plane. The transformation behavior for the components parallel and perpendicular to the surface normal (z-direction) are listed in Tab.~\ref{table}.}
    \label{FIG2}
\end{figure}

\begin{table}[htbp]
    \centering
    \begin{tabular}{|l|l|l|} \hline
           & $\mathcal{P}=m_{xy}\otimes m_{yz}\otimes m_{zx}$   & $\mathcal{R}_x(\pi)=m_{xz} \otimes m_{xy}$ \\ \hline \hline
         Chirality  & \textcolor{red}{$\chi \rightarrow -\chi$}& \textcolor{blue}{$\chi \rightarrow \chi$} \\ 
         D-vector& \textcolor{red}{$\textbf{D}_{||} \rightarrow -\textbf{D}_{||}$}& \textcolor{red}{$\textbf{D}_{||} \rightarrow -\textbf{D}_{||}$}\\
                 & \textcolor{blue}{$D_{\perp} \rightarrow D_{\perp}$}& \textcolor{red}{$D_{\perp} \rightarrow -D_{\perp}$}\\
         P-vector& \textcolor{blue}{$P_{||} \rightarrow P_{||}$} & \textcolor{red}{$P_{||} \rightarrow -P_{||}$}\\ 
                & \textcolor{red}{$P_{\perp} \rightarrow -P_{\perp}$}& \textcolor{blue}{$P_{\perp} \rightarrow P_{\perp}$}\\ \hline 
    \end{tabular}
    \caption{Symmetry behavior of the $\vec{D}$- and $\vec{P}$-vector under $\mathcal{P}$ and $\mathcal{R}$ operation  expressed as a product of successive orthogonal mirror operation $m$.  The parallel (||) and perpendicular ($\perp$) components of the vectors refer to the surface normal which coincides the mirror line (plane). The even and odd behavior is marked in blue and red, respectively. Only the $D_{||}$-part (marked in bold) of the molecular orbital moment changes sign for both symmetry operations. \label{table}}
\end{table}

In Fig.~\ref{FIG2}, we illustrate how the $\vD$ and $\vP$ vectors are changing  under a rotation of 180$^\circ$ and under a mirror reflection. We use a single alanine molecule in its zwitterionic form as a close analogue of the actual PA employed in the experiments. The molecule has charged groups at each end, creating an electric dipole, and it is chiral, existing in both L- and D- enantiomers. The details of the two vectors are summarized in Tab.~\ref{table}. With respect to chirality, the mirror plane serves as the central symmetry element. For our experiments with the out-of-plane magnetization of the Co layer, the surface normal lies within the mirror plane. As it is evident from the transformation behavior, only the component of the molecular orbital along the surface normal changes sign under both the parity and rotation operations, in agreement with our experimental findings.

%\section*{Conclusion}
\noindent\textit{Conclusions --} In the experimental part we perform conductance experiments on chiral molecules using an STM and magentic substrate. The orientation and the handedness of the molecules are carefully controlled. We find magnetoresistance, i.e.~a change of conductance, as the direction of the Co-layer magnetization is reversed. The results can be summarized by a term in the conductance of the form $\alpha \vmm\cdot\vm$. The $\vmm$ vector we interpret as a magnetization induced in the Au-layer by the presence of the chiral molecule. We suggest that this magnetization due to  not yet fully understood mechanism, is proportional to the magnetic dipole moment operator of the molecule, $\vL$. This enters the SOC in the form $\vL\cdot\vs$. 

This hypothesis not only explains the experimental findings that we present here, but also similar experiments by Aragonés et al.\cite{Aragones2025} Furthermore, recently reported negative CISS experiments \cite{Latha2025} are explained by noting that they involve an average of randomly oriented molecules. 
Finally, the hypothesis also help explain the correlation between the magnetization direction of nanomagnets with adsorbed chiral molecules.\cite{Koplovitz2019} Here the entire magnetization of the nanomagnet plays the role of $\vm$. When the adsorbed molecules are predominantly aligned parallel to one another on the magnet the $\vm$ will follow that common orientation. In the case  when the molecules have the different orientations, the individual contributions averages out. Therefore nanomagnets covered by molecules are magnetically indistinguishable from uncovered nanomagnets. 

We hope this hypothesis will instigate the effort in establishing  the correct theory for the intriguing CISS effect. 

\section*{Acknowledgements}
Funded by the Deutsche Forschungsgescheinschaft (DFG, German Research Foundation) - TRR 386 - A04 and Z02 (514664767).

\bibliographystyle{apsrev4-2}
\bibliography{Lit_CISS}

@article{london1937theorie,
  title={Th{\'e}orie quantique des courants interatomiques dans les combinaisons aromatiques},
  author={London, Fritz},
  journal={Journal de Physique et le Radium},
  volume={8},
  number={10},
  pages={397--409},
  year={1937},
  publisher={Soci{\'e}t{\'e} Fran{\c{c}}aise de Physique},
  doi={10.1051/jphysrad:01937008010039700}
}

@article{ditchfield1974self,
  title={Self-consistent perturbation theory of diamagnetism: I. A gauge-invariant LCAO method for NMR chemical shifts},
  author={Ditchfield, Robert},
  journal={Molecular Physics},
  volume={27},
  number={4},
  pages={789--807},
  year={1974},
  publisher={Taylor \& Francis},
  doi={10.1080/00268977400100711}
}

@article{wolinski1990efficient,
  title={Efficient implementation of the gauge-independent atomic orbital method for NMR chemical shift calculations},
  author={Wolinski, Krzysztof and Hinton, James F and Pulay, Peter},
  journal={Journal of the American Chemical Society},
  volume={112},
  number={23},
  pages={8251--8260},
  year={1990},
  publisher={ACS Publications},
  doi={10.1021/ja00179a005}
}

@article{Baljozovic2026,
author = {Baljozović, Miloš and Karmakar, Shiladitya and Fernandes Cauduro, Andr{\'e} L. and Shyam Sundar, Mothuku and Lozano, Marco and Kumar, Manish and Soler-Polo, Diego and Schmid, Andreas K. and Bedekar, Ashutosh V. and Jelinek, Pavel and Ernst, Karl-Heinz},
title = {Adsorption-Induced Surface Magnetism},
journal = {ACS Nano},
volume = {20},
number = {5},
pages = {4143-4151},
year = {2026},
doi = {10.1021/acsnano.5c15791},
    note ={PMID: 41614566},

URL = { 
    
        https://doi.org/10.1021/acsnano.5c15791
    
    

},
eprint = { 
    
        https://doi.org/10.1021/acsnano.5c15791
    
    

}
}

@article{Koplovitz2019,
Author = {Koplovitz, Guy and Leitus, Gregory and Ghosh, Supriya and Bloom, Brian
   P. and Yochelis, Shira and Rotem, Dvir and Vischio, Fabio and Striccoli,
   Marinella and Fanizza, Elisabetta and Naaman, Ron and Waldeck, David H.
   and Porath, Danny and Paltiel, Yossi},
Title = {Single Domain 10 nm Ferromagnetism Imprinted on Superparamagnetic
   Nanoparticles Using Chiral Molecules},
Journal = {SMALL},
Year = {2019},
Volume = {15},
Number = {1},
Month = {JAN 4},
Publisher = {WILEY-V C H VERLAG GMBH},
Address = {POSTFACH 101161, 69451 WEINHEIM, GERMANY},
Type = {Article},
Language = {English},
Affiliation = {Paltiel, Y (Corresponding Author), Hebrew Univ Jerusalem, Dept Appl Phys, IL-91904 Jerusalem, Israel.
   Porath, D; Paltiel, Y (Corresponding Author), Hebrew Univ Jerusalem, Ctr Nanosci \& Nanotechnol, IL-91904 Jerusalem, Israel.
   Porath, D (Corresponding Author), Hebrew Univ Jerusalem, Inst Chem, IL-91904 Jerusalem, Israel.
   Koplovitz, Guy; Yochelis, Shira; Paltiel, Yossi, Hebrew Univ Jerusalem, Dept Appl Phys, IL-91904 Jerusalem, Israel.
   Koplovitz, Guy; Yochelis, Shira; Rotem, Dvir; Porath, Danny; Paltiel, Yossi, Hebrew Univ Jerusalem, Ctr Nanosci \& Nanotechnol, IL-91904 Jerusalem, Israel.
   Koplovitz, Guy; Rotem, Dvir; Porath, Danny, Hebrew Univ Jerusalem, Inst Chem, IL-91904 Jerusalem, Israel.
   Leitus, Gregory, Weizmann Inst Sci, Dept Chem Res Support, IL-76100 Rehovot, Israel.
   Ghosh, Supriya; Bloom, Brian P.; Waldeck, David H., Univ Pittsburgh, Dept Chem, Pittsburgh, PA 15260 USA.
   Vischio, Fabio; Striccoli, Marinella, Natl Council Res CNR, Inst Chem \& Phys Proc IPCF, Via Orabona 4, I-70126 Bari, Italy.
   Fanizza, Elisabetta, Univ Bari, Dept Chem, Via Orabona 4, I-70126 Bari, Italy.
   Naaman, Ron, Weizmann Inst Sci, Dept Chem \& Biol Phys, IL-76100 Rehovot, Israel.},
DOI = {10.1002/smll.201804557},
Article-Number = {1804557},
ISSN = {1613-6810},
EISSN = {1613-6829},
Keywords-Plus = {RANDOM-ACCESS MEMORY; SPIN SELECTIVITY; SPINTRONICS},
Research-Areas = {Chemistry; Science \& Technology - Other Topics; Materials Science;
   Physics},
Web-of-Science-Categories  = {Chemistry, Multidisciplinary; Chemistry, Physical; Nanoscience \&
   Nanotechnology; Materials Science, Multidisciplinary; Physics, Applied;
   Physics, Condensed Matter},
Author-Email = {Danny.Porath@mail.huji.ac.il
   Paltiel@mail.huji.ac.il},
Affiliations = {Hebrew University of Jerusalem; Hebrew University of Jerusalem; Hebrew
   University of Jerusalem; Weizmann Institute of Science; Pennsylvania
   Commonwealth System of Higher Education (PCSHE); University of
   Pittsburgh; Consiglio Nazionale delle Ricerche (CNR); Istituto per i
   Processi Chimico-Fisici (IPCF-CNR); Universita degli Studi di Bari Aldo
   Moro; Weizmann Institute of Science},
ResearcherID-Numbers = {Paltiel, Yossi/AGQ-3659-2022
   Leitus, Gregory/K-1771-2012
   Vischio, Fabio/IZP-6355-2023
   Striccoli, Marinella/C-5456-2009
   Rotem, Dvir/G-1396-2013},
ORCID-Numbers = {Ghosh, Supriya/0000-0001-5088-2433
   Bloom, Brian/0000-0001-9581-9710
   Paltiel, Yossi/0000-0002-8739-9952
   Waldeck, David/0000-0003-2982-0929
   Leitus, Gregory/0000-0002-6749-6667
   Vischio, Fabio/0009-0002-0372-7523
   Striccoli, Marinella/0000-0002-5366-691X
   Rotem, Dvir/0000-0002-1840-6530},
Funding-Acknowledgement = {Volkswagen Foundation {[}VW 88 367]; Israel Science Foundation (ISF)
   {[}1589/14, 1248/10]; MOS Israel; European Research Council under the
   European Union's Seventh Framework Program (No. FP7/2007-2013)/ERC Grant
   {[}338720]; VW Foundation {[}VW 88 367]; U.S. Department of Energy
   {[}ER46430]; European Research Council (ERC) {[}338720] Funding Source:
   European Research Council (ERC)},
Funding-Text = {Y.P. acknowledges the support from the Volkswagen Foundation (No. VW 88
   367), the Israel Science Foundation (ISF Grant No. 1248/10), and the MOS
   Israel. R.N. acknowledges support in part from the European Research
   Council under the European Union's Seventh Framework Program (No.
   FP7/2007-2013)/ERC Grant Agreement No. (338720), the MOS Israel, and the
   VW Foundation (No. VW 88 367). D.P. acknowledges the Israel Science
   Foundation (ISF grant 1589/14) and the Minerva Centre for bio-hybrid
   complex systems. D.P. also thanks the Etta and Paul Schankerman Chair of
   Molecular Biomedicine. D.H.W. and R.N. acknowledge support by the U.S.
   Department of Energy; Grant No. ER46430.},
Number-of-Cited-References = {27},
Times-Cited = {51},
Usage-Count-Last-180-days = {2},
Usage-Count-Since-2013 = {67},
Journal-ISO = {Small},
Doc-Delivery-Number = {HG4OD},
Web-of-Science-Index = {Science Citation Index Expanded (SCI-EXPANDED)},
Unique-ID = {WOS:000454953900017},
OA = {Green Submitted},
DA = {2026-05-04},
}

@article{Latha2025,
author = {Li, Liang and Shi, Wanzhuo and Mahajan, Ankit and Zhang, Junxiang and Gómez-Gómez, Marta and Labella, Jorge and Louie, Shayan and Torres, Tomás and Barlow, Stephen and Marder, Seth R. and Reichman, David R. and Venkataraman, Latha},
title = {Too Fast for Spin Flipping: Absence of Chirality-Induced Spin Selectivity in Coherent Electron Transport through Single-Molecule Junctions},
journal = {Journal of the American Chemical Society},
volume = {147},
number = {28},
pages = {25043-25051},
year = {2025},
doi = {10.1021/jacs.5c08517},
    note ={PMID: 40601876},

URL = { 
    
        https://doi.org/10.1021/jacs.5c08517
    
    

},
eprint = { 
    
        https://doi.org/10.1021/jacs.5c08517
    
    

}

}

@article{Ha2022,
	author = {Nguyen, Thi Ngoc Ha and Rasabathina, Lokesh and Hellwig, Olav and Sharma, Apoorva and Salvan, Georgeta and Yochelis, Shira and Paltiel, Yossi and Baczewski, Lech T. and Tegenkamp, Christoph},
	date = {2022/08/24},
	doi = {10.1021/acsami.2c08668},
	isbn = {1944-8244},
	journal = {ACS Applied Materials \& Interfaces},
	journal1 = {ACS Applied Materials \& Interfaces},
	journal2 = {ACS Appl. Mater. Interfaces},
	month = {08},
	number = {33},
	pages = {38013--38020},
	publisher = {American Chemical Society},
	title = {Cooperative Effect of Electron Spin Polarization in Chiral Molecules Studied with Non-Spin-Polarized Scanning Tunneling Microscopy},
	type = {doi: 10.1021/acsami.2c08668},
	url = {https://doi.org/10.1021/acsami.2c08668},
	volume = {14},
	year = {2022},
	year1 = {2022}}

@article{Ha2023,
	author = {Ha Nguyen, Thi Ngoc and Paltiel, Yossi and Baczewski, Lech T. and Tegenkamp, Christoph},
	date = {2023/04/05},
	doi = {10.1021/acsami.3c01429},
	isbn = {1944-8244},
	journal = {ACS Applied Materials \& Interfaces},
	journal1 = {ACS Applied Materials \& Interfaces},
	journal2 = {ACS Appl. Mater. Interfaces},
	month = {04},
	number = {13},
	pages = {17406--17412},
	publisher = {American Chemical Society},
	title = {Spin Polarization of Polyalanine Molecules in 2D and Dimer-Row Assemblies Adsorbed on Magnetic Substrates: The Role of Coupling, Chirality, and Coordination},
	type = {doi: 10.1021/acsami.3c01429},
	url = {https://doi.org/10.1021/acsami.3c01429},
	volume = {15},
	year = {2023},
	year1 = {2023}}

@article{Gohler2011,
	author = {G{\"o}hler B. and Hamelbeck V. and Markus T. Z. and Kettner M. and Hanne G. F. and Vager Z. and Naaman R. and Zacharias H.},
	date = {2011/02/18},
	doi = {10.1126/science.1199339},
	journal = {Science},
	journal1 = {Science},
	journal2 = {Science},
	month = {2021/12/22},
	number = {6019},
	pages = {894--897},
	publisher = {American Association for the Advancement of Science},
	title = {{Spin Selectivity in Electron Transmission Through Self-Assembled Monolayers of Double-Stranded DNA}},
	type = {doi: 10.1126/science.1199339},
	url = {https://doi.org/10.1126/science.1199339},
	volume = {331},
	year = {2011},
	year1 = {2011}}

@article{Ha2019,
	author = {Nguyen, T. N. Ha and Solonenko, D. and Selyshchev, O. and Vogt, P. and Zahn, D. R. T. and Yochelis, S. and Paltiel, Y. and Tegenkamp, C.},
	booktitle = {The Journal of Physical Chemistry C},
	da = {2019/01/10},
	date = {2019/01/10},
	doi = {10.1021/acs.jpcc.8b10620},
	isbn = {1932-7447},
	journal = {The Journal of Physical Chemistry C},
	journal1 = {J. Phys. Chem. C},
	m3 = {doi: 10.1021/acs.jpcc.8b10620},
	month = {01},
	number = {1},
	pages = {612--617},
	publisher = {American Chemical Society},
	title = {{Helical Ordering of $\alpha$-l-Polyalanine Molecular Layers By Interdigitation}},
	ty = {JOUR},
	url = {https://doi.org/10.1021/acs.jpcc.8b10620},
	volume = {123},
	year = {2019},
	year1 = {2019}}

@article{Naaman2019,
	author = {Naaman, Ron and Paltiel, Yossi and Waldeck, David H.},
	da = {2019/04/01},
	doi = {10.1038/s41570-019-0087-1},
	id = {Naaman2019},
	isbn = {2397-3358},
	journal = {Nature Reviews Chemistry},
	number = {4},
	pages = {250--260},
	title = {{Chiral Molecules and The Electron Spin}},
	ty = {JOUR},
	url = {https://doi.org/10.1038/s41570-019-0087-1},
	volume = {3},
	year = {2019}}

@article{Kiran2016,
	author = {Kiran,Vankayala and Cohen,Sidney R. and Naaman,Ron},
	booktitle = {The Journal of Chemical Physics},
	da = {2017/03/07},
	date = {2016/11/18},
	doi = {10.1063/1.4966237},
	isbn = {0021-9606},
	journal = {The Journal of Chemical Physics},
	journal1 = {J. Chem. Phys.},
	m3 = {doi: 10.1063/1.4966237},
	month = {2021/12/15},
	number = {9},
	pages = {092302},
	publisher = {American Institute of Physics},
	title = {{Structure Dependent Spin Selectivity in Electron Transport Through Oligopeptides}},
	ty = {JOUR},
	url = {https://doi.org/10.1063/1.4966237},
	volume = {146},
	year = {2016},
	year1 = {2016}}

@article{Mondal2021,
	author = {Mondal, Amit Kumar and Preuss, Marco D. and Sleczkowski, Marcin L. and Das, Tapan Kumar and Vantomme, Ghislaine and Meijer, E. W. and Naaman, Ron},
	comment = {doi: 10.1021/jacs.1c02983},
	doi = {10.1021/jacs.1c02983},
	journal = {J. Am. Chem. Soc.},
	number = {18},
	pages = {7189},
	publisher = {American Chemical Society},
	title = {{Spin Filtering in Supramolecular Polymers Assembled from Achiral Monomers Mediated by Chiral Solvents}},
	url = {https://doi.org/10.1021/jacs.1c02983},
	volume = {143},
	year = {2021}}

@Article{Ha2020,
  author    = {Ha, Nguyen T. N. and Sharma, A. and Slawig, D. and Yochelis, S. and Paltiel, Y. and Zahn, D. R. T. and Salvan, G. and Tegenkamp, C.},
  journal   = {J. Phys. Chem. C},
  title     = {{Charge-Ordered $\alpha$-Helical Polypeptide Monolayers on Au(111)}},
  year      = {2020},
  issn      = {1932-7447},
  month     = mar,
  number    = {10},
  pages     = {5734--5739},
  volume    = {124},
  comment   = {doi: 10.1021/acs.jpcc.0c00246},
  doi       = {10.1021/acs.jpcc.0c00246},
  publisher = {American Chemical Society},
  url       = {https://doi.org/10.1021/acs.jpcc.0c00246},
}

@Article{Tirion2024,
  author    = {Tirion, Sytze H. and van Wees, Bart J.},
  journal   = {ACS Nano},
  title     = {Mechanism for Electrostatically Generated Magnetoresistance in Chiral Systems without Spin-Dependent Transport},
  year      = {2024},
  issn      = {1936-0851},
  month     = feb,
  number    = {8},
  pages     = {6028--6037},
  volume    = {18},
  comment   = {doi: 10.1021/acsnano.3c12925},
  doi       = {10.1021/acsnano.3c12925},
  publisher = {American Chemical Society},
  url       = {https://doi.org/10.1021/acsnano.3c12925},
}

@Article{Yang2019,
  author    = {Yang, Xu and van der Wal, Caspar H. and van Wees, Bart J.},
  journal   = {Phys. Rev. B},
  title     = {Spin-dependent electron transmission model for chiral molecules in mesoscopic devices},
  year      = {2019},
  month     = {Jan},
  pages     = {024418},
  volume    = {99},
  doi       = {10.1103/PhysRevB.99.024418},
  issue     = {2},
  numpages  = {16},
  publisher = {American Physical Society},
  url       = {https://link.aps.org/doi/10.1103/PhysRevB.99.024418},
}

@Article{Clever2022,
  author    = {Clever, Caleb and Wierzbinski, Emil and Bloom, Brian P. and Lu, Yiyang and Grimm, Haley M. and Rao, Shilpa R. and Horne, W. Seth and Waldeck, David H.},
  journal   = {Isr. J. Chem.},
  title     = {Benchmarking Chiral Induced Spin Selectivity Measurements - Towards Meaningful Comparisons of Chiral Biomolecule Spin Polarizations},
  year      = {2022},
  issn      = {0021-2148},
  month     = dec,
  number    = {11-12},
  pages     = {e202200045},
  volume    = {62},
  doi       = {10.1002/ijch.202200045},
  publisher = {John Wiley & Sons, Ltd},
  url       = {https://doi.org/10.1002/ijch.202200045},
}

@Article{EckshtainLevi2016,
  author   = {Eckshtain-Levi, Meital and Capua, Eyal and Refaely-Abramson, Sivan and Sarkar, Soumyajit and Gavrilov, Yulian and Mathew, Shinto P. and Paltiel, Yossi and Levy, Yaakov and Kronik, Leeor and Naaman, Ron},
  journal  = {Nature Communications},
  title    = {Cold denaturation induces inversion of dipole and spin transfer in chiral peptide monolayers},
  year     = {2016},
  issn     = {2041-1723},
  number   = {1},
  pages    = {10744},
  volume   = {7},
  doi      = {10.1038/ncomms10744},
  refid    = {Eckshtain-Levi2016},
  url      = {https://doi.org/10.1038/ncomms10744},
}

@Article{Eckvahl2023,
  author    = {Eckvahl, Hannah J. and Tcyrulnikov, Nikolai A. and Chiesa, Alessandro and Bradley, Jillian M. and Young, Ryan M. and Carretta, Stefano and Krzyaniak, Matthew D. and Wasielewski, Michael R.},
  journal   = {Science},
  title     = {Direct observation of chirality-induced spin selectivity in electron donor–acceptor molecules},
  year      = {2023},
  issn      = {1095-9203},
  month     = oct,
  number    = {6667},
  pages     = {197--201},
  volume    = {382},
  doi       = {10.1126/science.adj5328},
  publisher = {American Association for the Advancement of Science (AAAS)},
}

@Article{Ha2024,
author ="Nguyen, Thi Ngoc Ha and Salvan, Georgeta and Hellwig, Olav and Paltiel, Yossi and Baczewski, Lech Thomasz and Tegenkamp, Christoph",
title  ="The mechanism of the molecular CISS effect in chiral nano-junctions",
journal  ="Chem. Sci.",
year  ="2024",
volume  ="15",
issue  ="36",
pages  ="14905-14912",
publisher  ="The Royal Society of Chemistry",
doi  ="10.1039/D4SC04435E",
url  ="http://dx.doi.org/10.1039/D4SC04435E"}

@Article{Aragones2025,
  author    = {Aragonès, Albert C. and Varese, Monica and Garg, Kavita and Kuang, Wenzhu and Wang, Qiankun and Giralt, Ernest and Mujica, Vladimiro and Gutierrez, Rafael and Cuniberti, Gianaurelio and Puerta, Luis and Díez-Pérez, Ismael},
  journal   = {J. Am. Chem. Soc.},
  title     = {Dipole-Induced Inversion of Spin-Dependent Charge Transport through α-Helical Peptide-Based Single-Molecule Junctions},
  year      = {2025},
  issn      = {0002-7863},
  month     = oct,
  number    = {40},
  pages     = {36453--36463},
  volume    = {147},
  comment   = {doi: 10.1021/jacs.5c10892},
  doi       = {10.1021/jacs.5c10892},
  publisher = {American Chemical Society},
  url       = {https://doi.org/10.1021/jacs.5c10892},
}

@Article{Rana2025,
  author    = {Rana, Shammi and Remigio, Massimiliano and Aravindan Geetha, Lekshmi and Strutyński, Karol and Volpi, Martina and John, Sanjay and Baczewski, Lech Tomasz and Paltiel, Yossi and Resel, Roland and Melle-Franco, Manuel and Mali, Kunal S. and Geerts, Yves H. and De Feyter, Steven},
  title     = {Chirality-Induced Spin Selectivity in Two-Dimensional Self-Assembled Molecular Networks},
  doi       = {10.1021/jacs.5c12143},
  issn      = {0002-7863},
  number    = {46},
  pages     = {42426--42432},
  url       = {https://doi.org/10.1021/jacs.5c12143},
  volume    = {147},
  comment   = {doi: 10.1021/jacs.5c12143},
  journal   = {J. Am. Chem. Soc.},
  month     = nov,
  publisher = {American Chemical Society},
  year      = {2025},
}

@Article{SM,
  author    = {Nguyen et al.},
  journal   = {Supplemental Material},
  title     = {Supplemental Material},
  year      = {2026},
  issn      = {},
  number    = {},
  pages     = {},
  volume    = {},
  doi       = {XXX},
  publisher = {American Chemical Society},
  url       = {},
}

@Article{Mishra2013,
  author    = {Mishra, Debabrata and Markus, Tal Z. and Naaman, Ron and Kettner, Matthias and Göhler, Benjamin and Zacharias, Helmut and Friedman, Noga and Sheves, Mordechai and Fontanesi, Claudio},
  title     = {Spin-dependent electron transmission through bacteriorhodopsin embedded in purple membrane},
  doi       = {10.1073/pnas.1311493110},
  number    = {37},
  pages     = {14872--14876},
  url       = {https://doi.org/10.1073/pnas.1311493110},
  volume    = {110},
  comment   = {doi: 10.1073/pnas.1311493110},
  journal   = {Proceedings of the National Academy of Sciences},
  month     = sep,
  publisher = {Proceedings of the National Academy of Sciences},
  year      = {2013},
}

@Article{Amsallem2023,
  author       = {Amsallem, Dana and Kumar, Anil and Naaman, Ron and Gidron, Ori},
  year         = {2023},
  month        = {sep},
  journal      = {Chirality},
  title        = {Spin polarization through axially chiral linkers: Length dependence and correlation with the dissymmetry factor},
  doi          = {10.1002/chir.23556},
  issn         = {0899-0042},
  number       = {9},
  pages        = {562--568},
  url          = {https://doi.org/10.1002/chir.23556},
  volume       = {35},
  publisher    = {John Wiley & Sons, Ltd},
}

@Article{Bloom2024,
  author    = {Bloom, Brian P. and Paltiel, Yossi and Naaman, Ron and Waldeck, David H.},
  title     = {Chiral Induced Spin Selectivity},
  doi       = {10.1021/acs.chemrev.3c00661},
  issn      = {0009-2665},
  number    = {4},
  pages     = {1950--1991},
  url       = {https://doi.org/10.1021/acs.chemrev.3c00661},
  volume    = {124},
  comment   = {doi: 10.1021/acs.chemrev.3c00661},
  journal   = {Chem. Rev.},
  month     = feb,
  publisher = {American Chemical Society},
  year      = {2024},
}

@Article{Ray1999,
  author       = {Ray, K. and Ananthavel, S. P. and Waldeck, D. H. and Naaman, R.},
  year         = {1999},
  month        = {feb},
  journal      = {Science},
  title        = {Asymmetric Scattering of Polarized Electrons by Organized Organic Films of Chiral Molecules},
  doi          = {10.1126/science.283.5403.814},
  number       = {5403},
  pages        = {814--816},
  url          = {https://doi.org/10.1126/science.283.5403.814},
  volume       = {283},
  comment      = {doi: 10.1126/science.283.5403.814},
  publisher    = {American Association for the Advancement of Science},
}

@ARTICLE{Buttiker1988,
  author={Buttiker, M.},
  journal={IBM Journal of Research and Development}, 
  title={Symmetry of electrical conduction}, 
  year={1988},
  volume={32},
  number={3},
  pages={317-334},
  doi={10.1147/rd.323.0317}}

@article{Buttiker1986,
  title = {Four-Terminal Phase-Coherent Conductance},
  author = {B\"uttiker, M.},
  journal = {Phys. Rev. Lett.},
  volume = {57},
  issue = {14},
  pages = {1761--1764},
  numpages = {0},
  year = {1986},
  month = {Oct},
  publisher = {American Physical Society},
  doi = {10.1103/PhysRevLett.57.1761},
  url = {https://link.aps.org/doi/10.1103/PhysRevLett.57.1761}
}

@Article{Schuhl2005,
  author       = {Schuhl, Alain and Lacour, Daniel},
  year         = {2005},
  journal      = {Comptes Rendus Physique},
  title        = {Spin dependent transport: GMR & TMR},
  doi          = {10.1016/j.crhy.2005.10.010},
  issn         = {1631-0705},
  number       = {9},
  pages        = {945--955},
  url          = {https://www.sciencedirect.com/science/article/pii/S1631070505001647},
  volume       = {6},
}

\end{document}

\title{Supplement Material: The geometry of the CISS effect}

\author{P. Hedegård}
\email{hedegard@nbi.ku.dk}
\affiliation{Niels Bohr Institute, University of Copenhagen, Denmark}

\author{A. Kazimir}
\author{C. Lamers}
\affiliation{Institute for Drug Discovery, Faculty of Medicine, Leipzig University, Germany}

\author{L.T. Baczewski}
\affiliation{Institute of Physics, Polish Academy of Sciences, Warszawa, Poland}

\author{T.N.H. Nguyen}
\affiliation{Institute of Physics, Chemnitz University of Technology, Germany}

\author{C. Tegenkamp}
\email{christoph.tegenkamp@physik.tu-chemnitz.de}
\affiliation{Institute of Physics, Chemnitz University of Technology, Germany}

\maketitle

\section{Structure and self-assembly process}

\begin{figure}[H]
        \centering
    \includegraphics[width=0.8\linewidth]{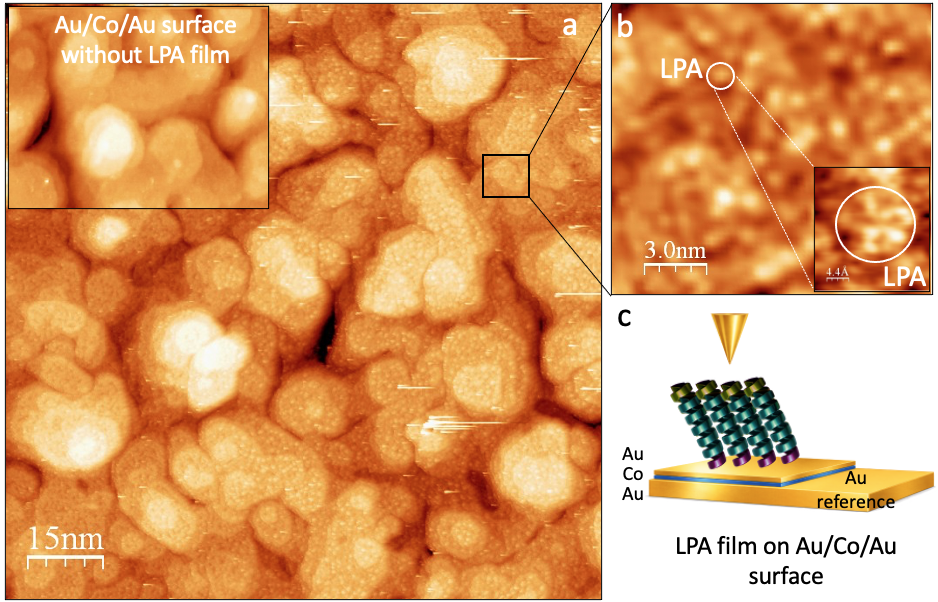}
    \caption{(a) STM image exemplarily for L-PA (Cys-N-terminus or Cys-C-terminus) molecules adsorbed on the Au/Co/Au surface. The inset in (a) shows the bare Au/Co/Au surface without the L-PA film. (b) Magnified view of the black rectangular area in (a). The insert in (b) shows a single L-PA molecule exhibiting a three-fold symmetry characteristic of the helical backbone of the PA peptide \cite{Ha2019,Ha2020}. (c) Schematic illustration of the L-PA film adsorbed on the Au/Co/Au surface and the Au surface reference.}
    \label{FIGS1}
\end{figure}

As magnetic substrates the epitaxially grown Au/Co/Au structures were used. Half of the surface was non-magnetic, i.e. revealing only Au. This allowed us to perform reference measurements without a magnetic layer as sketched in Fig.~\ref{FIGS1}c). 
Both Au surfaces show terrace sizes of $\sim$20~nm, sufficient for STM/STS measurements (cf. Fig.~\ref{FIGS1}a). The nanostructures posses a perpendicular anisotropy whar was confirmed by polar MOKE. 

In this study, 0.3 mM solutions of PA molecules in ethanol were used. The self-assembled structures were prepared by drop-casting. STM experiments were performed after the films were dried.
We first characterized the self-assembled films of the Cys-C-terminus and Cys-N-terminus L-PA peptides on the Au/Co/Au surface. Both films exhibited well-ordered, interdigitated structures, chemically bond on Au/Co/Au surfaces, as shown in the STM images in Figure 1, acquired at a tunneling current of 100 pA and a bias voltage of 0.5 V at 300K. Figure \ref{FIGS1}a) presents a large-area STM image (120 × 120 nm²) of the Au/Co/Au substrate covered with L-PA peptides; the inset shows the bare Au surface prior to peptide deposition. The surface roughness (~0.2 nm) and terrace sizes (~20 nm) are comparable for both surfaces (Au references and Au/Co/Au surface) and are suitable for high-quality STM and STS measurements. Figure \ref{FIGS1}b) shows a magnified region from Fig.~\ref{FIGS1}a), revealing that the L-PA peptides self-assemble into a highly ordered interdigitated structure over the surface. Each bright spot corresponds to a single L-PA molecule (white circle in Fig.~\ref{FIGS1}b). This observed self-assembled structure is consistent with our earlier reports \cite{Ha2019,Ha2020}.

\section{STS measurements for the determination of the CISS-MR}

\begin{figure}
        \includegraphics[width=0.5\linewidth]{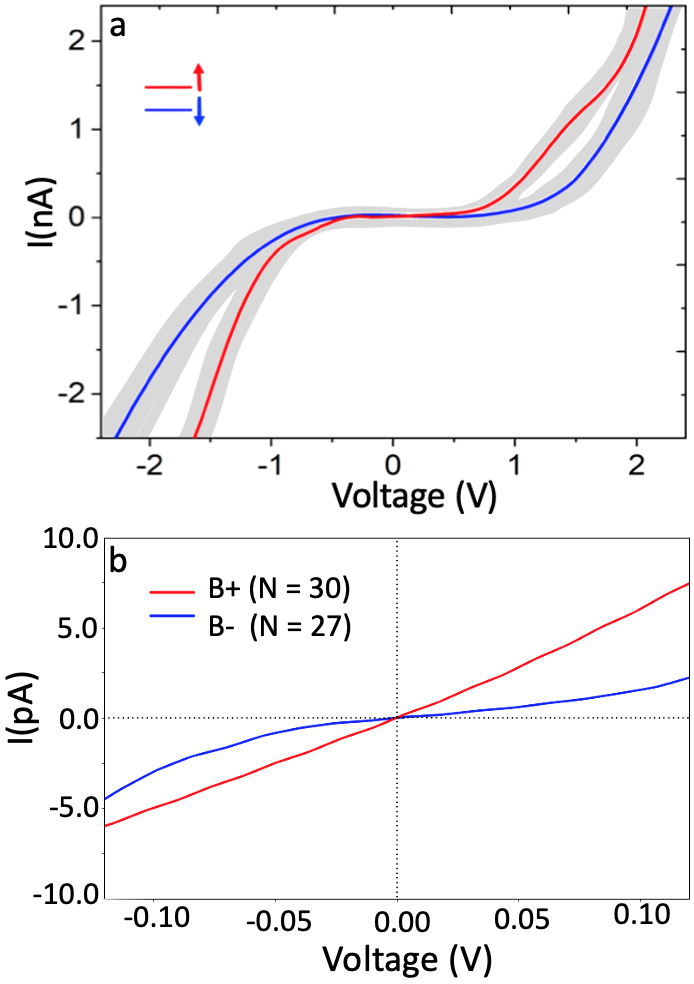}
        \caption{(a) STS measurements for the two magnetic configurations. the gray colored lines show the variation of the IV curves for our ambient setup. Nevertheless, the colored lines represent the averages and allow to determine the CISS-MR effect. (b) STS curves of a similar molecule on Au/Co/Au measured in ultra-high vacuum at 5~K showing also a CISS-MR effect at low biases.}
    \label{FIGS2}
\end{figure}

The CISS experiments presented in this paper were acquired following the procedure described in Ref.~\cite{Ha2024}. The variation of the I--V data obtained under ambient conditions is shown in Fig.~\ref{FIGS2}(a). The solid (colored) lines represent averaged data and clearly reveal the CISS-MR effect. Each IV curve is an average of at least ten measurements and the average was taken at least from 10 IV curves. 

In addition, test measurements were performed on 16-mer LPA molecules deposited on Au/Co/Au at 5~K under ultra-high-vacuum conditions Fig.~\ref{FIGS2}b). The molecules for these reference measurements were adsorbed ex situ by drop casting. Compared to the data obtained under ambient conditions, distinct differences in the I--V characteristics are observed at low bias voltages, i.e., in the linear regime.

\begin{figure}[H]
            \centering
            \includegraphics[width=0.5\linewidth]{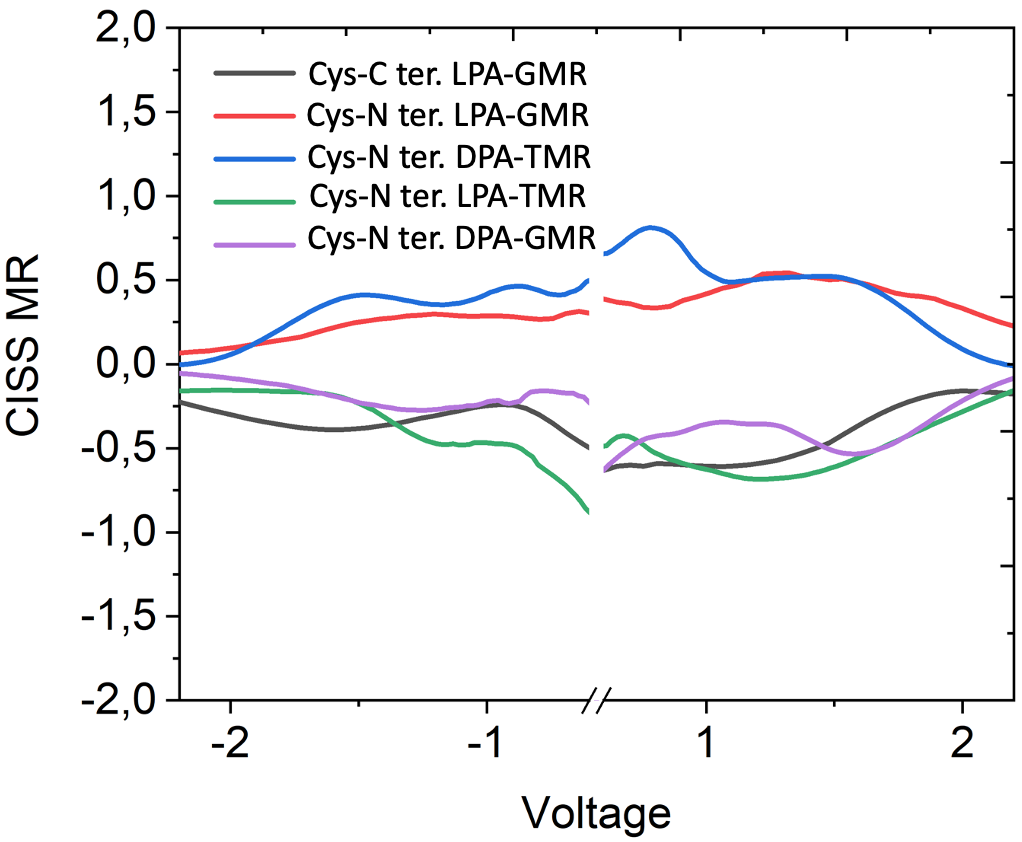}
        \caption{CISS magnetoresistance (CISS-MR) measured for both GMR and TMR hybrid systems shown in Fig. 2. A positive CISS-MR is observed for L-PA peptides with the dipole moment oriented downward toward the surface (red curve) and for D-PA peptides with the dipole moment oriented upward (purple curve). In contrast, a negative CISS-MR is obtained for N-terminus L-PA peptides attached to the Au tip with the dipole moment oriented upward (green curve), as well as for N-terminus D-PA peptides with the dipole moment pointing upward (blue curve) away from the surface.}
    \label{FIGS3}
\end{figure}

All corresponding results are summarized in Fig.~\ref{FIGS3}, which clearly demonstrates that both the molecular helicity and the orientation of the molecular dipole moment play decisive roles in determining the sign and magnitude of the spin polarization. 

For identical dipole orientations but opposite helicities, the left-handed peptide films exhibit a lower CISS-MR of approximately 5–10\% compared with the right-handed films (purple and red curves in Fig.~\ref{FIGS3}, respectively). This reduced response likely originates from imperfections introduced during the synthesis of the D-peptide, which can result in less ordered self-assembled monolayers (SAMs) of D-PA, as reported previously. These observations are in good agreement with earlier studies demonstrating that molecular chirality, in combination with the spin–orbit coupling of the substrate, induces an asymmetry of a few percent in the transmission probabilities for spin-up and spin-down electrons. Reversing the molecular dipole moment while preserving the helicity (C-terminus to N-terminus L-PA) leads to a sign reversal of the CISS-MR from positive to negative, as illustrated by the black and green curves in Fig. 3, respectively. A similar inversion is observed when the helicity is reversed (from right- to left-handed peptide) while maintaining the same dipole orientation, resulting in an opposite CISS-MR, as shown by the red and purple curves in Fig~\ref{FIGS3}, respectively. The green curve corresponds to the TMR hybrid system, in which N-terminus L-PA is attached to the Au STM tip. In this configuration, the helicity is preserved, but the dipole moment points upward toward the tip and away from the magnetic surface. The resulting CISS-MR is negative similar to the case with identical helicity and upward-oriented dipole (Cys-C ter. LPA GMR) in Figure 2a. Interestingly, the CISS-MR of the TMR hybrid system reaches values up to 15\% higher than that of the GMR hybrid for peptides with the same helicity and length. This difference reflects the role of the intrinsic electric field within the helical backbone of the peptide. The dipolar field dominates the local electrostatic environment, particularly when the peptide is attached to the STM tip, where the electric field between tip and substrate is highly inhomogeneous and strongest near the tip apex. This observation demonstrates the dominant role of the molecular dipole in controlling spin-polarized transport.

\section{Synthesis and characterization of alanine-rich peptides}

The alanine-rich helices of the sequence (right-handed, L-, Cys\textsubscript{N}-, left-handed D-, Cys\textsubscript{N}- 
) C[AAAAK]\textsubscript{3} were obtained from Sigma Aldrich. Polyalanine (PA) with the sequence (right-handed, L-, Cys\textsubscript{C}-) [AAAAK]\textsubscript{3}C was synthesized using a microwave-assisted solid-phase peptide synthesis strategy (SPPS).
 Amino acids and deprotection reagents were obtained from Sigma-Aldrich, coupling reagents were purchased from Carbolution, solvents (DMF and DCM) were obtained from VWR and used without further purification. 
Synthesis was performed on Fmoc-Cys(Trt)-Wang resin (loading 0.334 mmol/g). The resin was preloaded with cysteine protected at the thiol with a trityl (Trt) group. The synthesis was carried out on a microwave peptide synthesizer under standard automated protocol.
Fmoc deprotection was achieved using 20\% (v/v) 4-methylpiperidine in DMF at 75 °C for two consecutive cycles (30 seconds followed by 3 minutes). After each deprotection step, the resin was washed thoroughly with DMF.
Amino acid couplings were performed using Fmoc-protected amino acids (5.0 equiv relative to resin loading) activated with DIC and Oxyma (5.0 equiv each) in DMF. Coupling reactions were conducted at 75 °C for 5 min under microwave irradiation.
To minimize formation of truncated sequences, a capping step was performed after each coupling cycle using acetic anhydride solution (5\% v/v Ac\textsubscript{2}O in DMF). The resin was subsequently washed with DMF before proceeding to the next deprotection step.
After completion of chain assembly and final Fmoc removal, the resin was washed with DMF followed by DCM and dried in air.
The peptide was cleaved from the resin and side-chain protecting groups were simultaneously removed using a cleavage cocktail consisting of TFA/TIS/EDT/H\textsubscript{2}O (92.5:2.5:2.5:2.5, v/v/v/v) for 3–4 h at room temperature.
The cleavage mixture was filtered to remove the resin, and the crude peptide was precipitated by addition of cold diethyl ether (Et\textsubscript{2}O). The precipitated peptide was collected by centrifugation, washed with cold ether, and dried at air.
Purification was performed by preparative reverse-phase high-performance liquid chromatography (RP-HPLC) using a linear gradient of 5–60\% acetonitrile in water containing 0.1\% TFA.
Fractions corresponding to the desired 16-mer peptide were identified by mass spectrometry, lyophilized, and dried to afford the final product.

\begin{figure}
    \centering
    \includegraphics[width=1\linewidth]{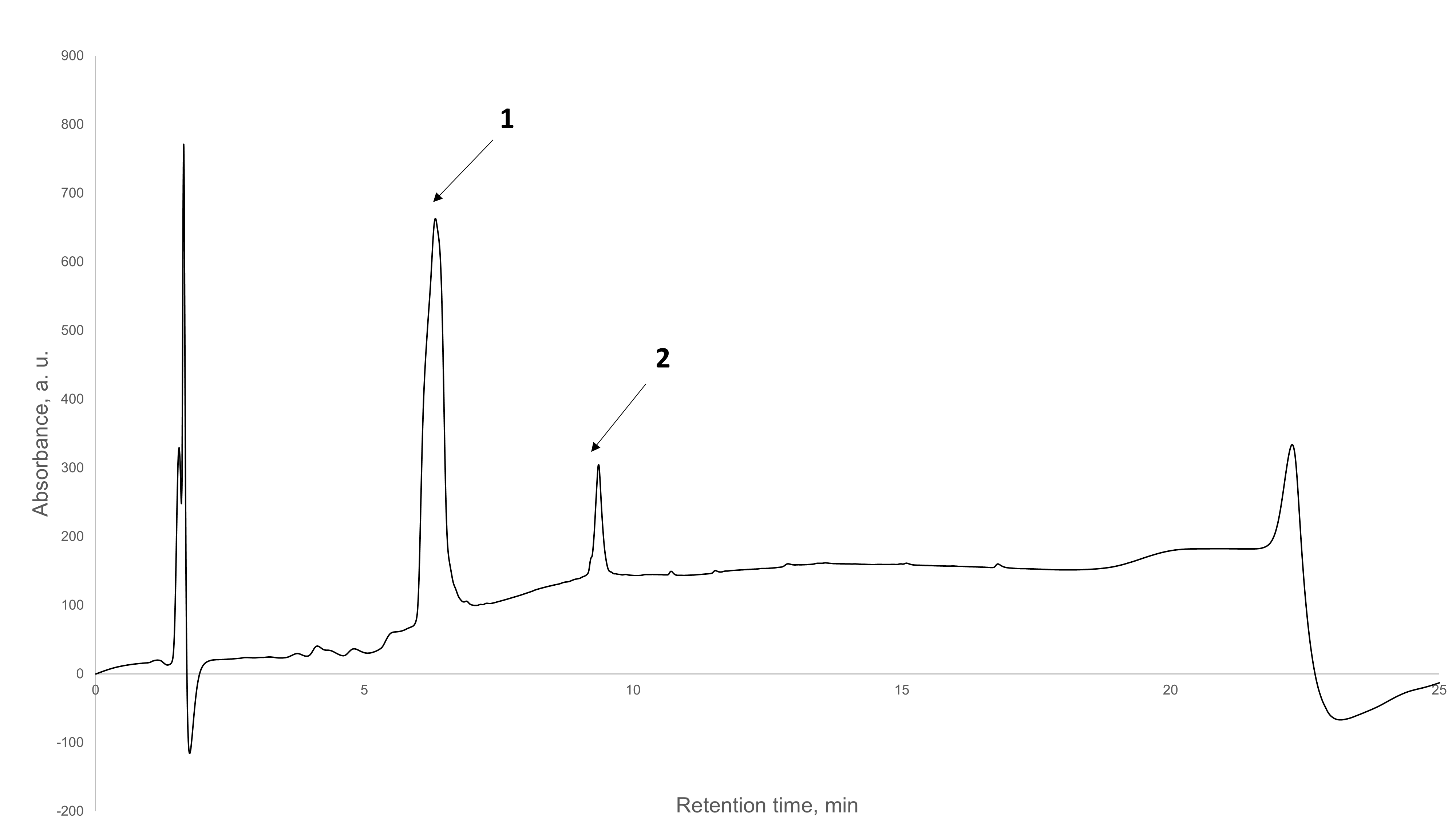}
\caption{RP-HPLC of [AAAAK]\textsubscript{3}C measured at 215 nm over 25 min with gradient 5 to 95\% of acetonitrile in water (+0.1\% of formic acid) using C18 column (see the parameters above) with the 0.40 ml/min at 40°C. Purity assessed was 89\% (1st peak belongs to the desired peptide).
}
\label{FIGS4}
 \end{figure}

\begin{figure}
     \centering
     \includegraphics[width=0.75\linewidth]{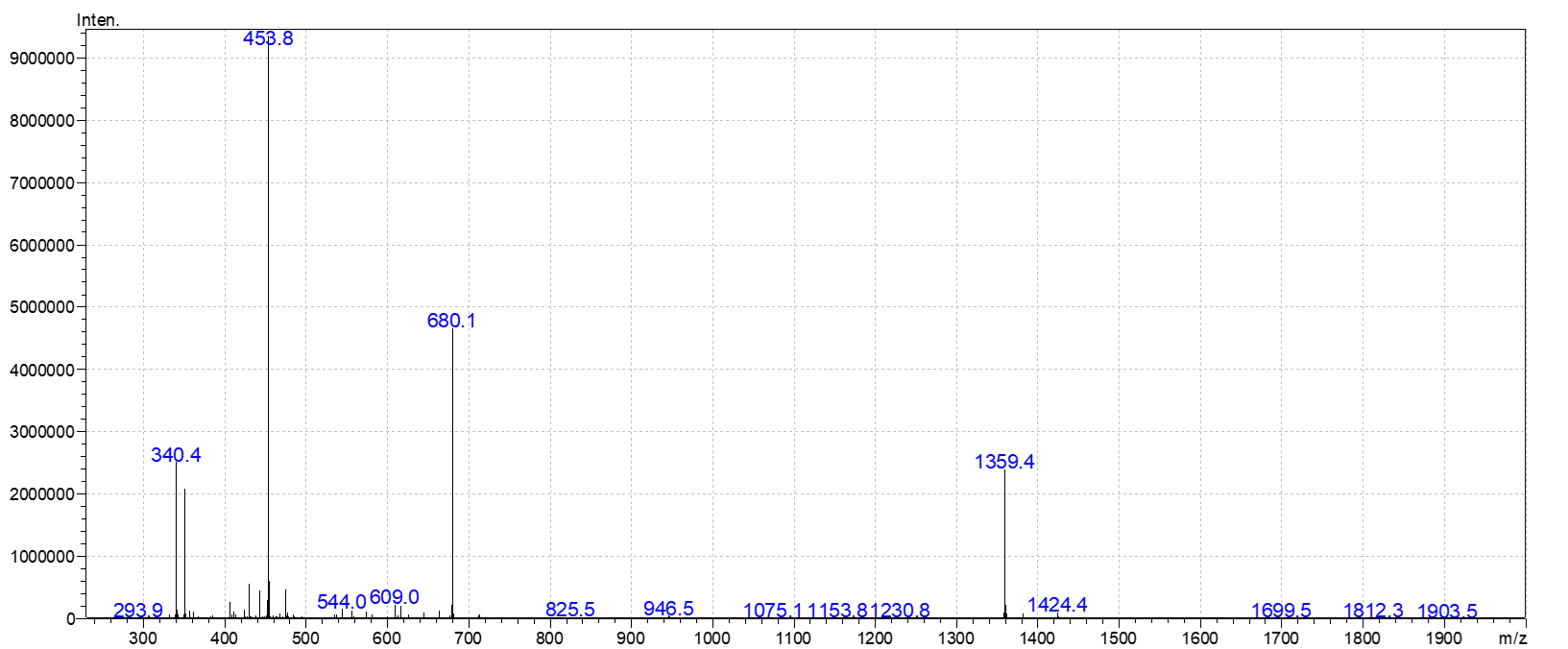}
     \caption{Mass spectrum of purified 16mer of the sequence [AAAAK]\textsubscript{3}C. M = 1358, masses (positive mode): [M+H]\textsuperscript{+}
   = 1359 m/z (calc), 1359 m/z (found),  [M+2H]\textsuperscript{2+}/2  = 680 m/z (calc), 680 m/z (found), [M+3H]\textsuperscript{3+}/3 = 454 m/z (calc), 454 m/z (found), [M+4H]\textsuperscript{4+}/4 = 340 m/z (calc), 340 m/z (found).   }
     \label{FIGS5}
 \end{figure}

Purity of the isolated peptide was determined to be 89\% by analytical RP-HPLC (cf. Fig.~\ref{FIGS4}) using a linear gradient of 5–95\% acetonitrile in water (both containing 0.1\% TFA) over 25 min. The HPLC data were obtained using Shimadzu Nexera LCMS 2020.  Analytical separations were carried out using an Atlantis T3 C18 column (150 × 2.1 mm, 3 $\mu$m particle size, 100 Å pore size) from Waters Corporation. A mass-spectrum of purified 16-mer of the the sequence [AAAAK]\textsubscript{3}C is shown in Fig.~\ref{FIGS5}.

\bibliographystyle{apsrev4-2}
\bibliography{Lit_CISS}